\documentclass{aa}  
\usepackage{amsmath}
\usepackage{graphicx}
\usepackage{appendix}
\usepackage{multirow}
\usepackage[varg]{txfonts}
\usepackage{natbib}
\usepackage{array}
\usepackage{nicefrac}
\bibpunct{(}{)}{;}{a}{}{,}
\usepackage[version=4]{mhchem}
\usepackage{soul}
\usepackage{threeparttable}
\usepackage[dvipsnames]{xcolor}
\usepackage{ulem}
\usepackage{comment}
\usepackage[breaklinks,pdfnewwindow=true,colorlinks=true,citecolor=blue]{hyperref}
\usepackage{tabularx}

\inputencoding{utf8}

\newcommand* \samethanks[1][\value{footnote}]{\footnotemark[#1]}

\begin{document}

\title{Broadband spectroscopy of astrophysical ice analogues}
\subtitle{V. Optical constants of
Ih, Ic, and amorphous \ce{H2O} ices
in the terahertz-infrared range}

\author{
    A.A.~Gavdush \inst{\ref{GPI}}\thanks{Both authors contributed equally to this work.},
    F.~Ribeiro \inst{\ref{MPE}, \ref{IFRJ}}\samethanks[1],
    M.K.~Matveishina \inst{\ref{GPI}},
     A.Vyjidak\inst{\ref{MPE}},
    F.~Kruczkiewicz \inst{\ref{MPE}, \ref{AMU}}\thanks{Present address: Leiden Observatory, Leiden University, Niels Bohrweg 2, 2333 CA Leiden, The Netherlands},
    G.A.~Komandin \inst{\ref{GPI}},
    S.V.~Garnov \inst{\ref{GPI}},
    K.I.~Zaytsev \inst{\ref{GPI}},
    A.V.~Ivlev \inst{\ref{MPE}},
    T.~Grassi \inst{\ref{MPE}},
    B.M.~Giuliano\inst{\ref{MPE}},
    and P.~Caselli \inst{\ref{MPE}}
}

\institute{Prokhorov General Physics Institute of
the Russian Academy of Sciences, 
119991 Moscow, Russia
\label{GPI}
\and
Max-Planck-Institut f\"ur Extraterrestrische Physik,
Gießenbachstraße 1, Garching, 85748, Germany
\label{MPE}
\and
Federal Institute of Education, Science and Technology of Rio de Janeiro (IFRJ), Nilópolis Campus - Rio de Janeiro - Brazil
\label{IFRJ}
\and 
Aix Marseille Univ, CNRS, CNES, LAM, Marseille, France 
\label{AMU}}

\date{Received - , 2024;
Accepted - , 2024}
   
\authorrunning{A.A.~Gavdush~et~al.}

\titlerunning{Optical constants of
Ih, Ic, and amorphous \ce{H2O} ices
in the THz and IR ranges}

\date{Received 2025; accepted  2025}
    
\abstract
    {Knowledge of the terahertz (THz)--infrared (IR) optical properties of astrophysical ices is important for understanding the dust continuum emission and radiative transfer in dense and cold interstellar environments. Water ice plays a dominant role in the aforementioned phenomena due to its prevalence and the high dipole moment of the \ce{H2O} molecule, resulting in high absorptivity and emissivity. Direct measurements of optical constants in the THz spectral region for astrophysically relevant \ce{H2O} ice samples are scarce. Extrapolation of optical properties in the THz spectral region from IR data can introduce uncertainties into astrophysical models.}
    {We measured the optical properties of water ice samples in the Ih and Ic forms as well as amorphous solid water (ASW) in the THz region in order to derive broad optical constants using literature and experimental data in the THz--IR range.}
    {In our experiments, the Ih, Ic, and ASW ices were grown by vapour deposition onto a cold substrate and measured by THz pulsed spectroscopy. Their THz optical properties were retrieved, compared with the THz--IR literature data, and approximated using the multiple-Lorentz model.}
    {From the existing literature data on the Ih, Ic, and ASW ices, we selected samples with the highest optical constants and classified them as compact. Their optical properties were merged in the frequency range of $\nu = 0.3$--$120$~THz (the wavelength range of $\lambda = 1$~mm--2.5$~\mu$m). The underlying absorption bands were attributed to vibrational modes and approximated using the multiple-Lorentz model while accounting for anharmonicity. 
    Discrepancies primarily arising in low-absorption regions between the experimental data and broadband models were attributed to factors such as the model's complexity and the baseline-subtraction procedure.
    The THz response of all ices is formed by the low-frequency wings of the IR bands and the single broad low-intense THz peak around $1.8$~THz, which is very similar for all phases. The opacity calculation for dust grains covered by \ce{H2O} ice mantles based on experimental data shows discrepancies with data derived by extrapolation.}
    {The inferred THz--IR optical constants of water ice are important for future observations and modelling of cold clouds and protoplanetary disks.}
    
\maketitle
\nolinenumbers

\section{Introduction}

Dust is a minor but key component of the interstellar medium (ISM), playing a crucial role in its physics and chemistry and particularly in ice formation by providing surfaces for ice-mantle growth. Dust also serves as an important tracer of overall gas abundance, as direct observation of molecular hydrogen is challenging \citep{Tielens_2013, Walsh2014, Bohlin_1978, Lombardi_2014}. The use of dust continuum emission to measure dust masses is particularly important in dense and cold regions, where other tracers of molecular gas, such as CO, are depleted \citep{Dutrey_1998, Caselli_1999}.

The dust grain size distribution and molecular composition, among other parameters, influence the optical properties and opacities of dust \citep{Li_2001, Weingartner_2001}, which can be further affected by the presence of ice mantles. Facilities such as the Atacama Large Millimeter/submillimeter Array (ALMA) and the Northern Extended Millimeter Array (NOEMA) provide observations of the dust continuum emission in the millimetre and submillimetre region and are the most powerful tools that can be used to investigate the characteristics of prestellar cores, young stars, and protoplanetary disks \citep[e.g.][]{Dullemond_2018, Ohashi2018, Andrews2018, Caselli2019, Sabatini_2025}.

Water is the most abundant component of ice mantles \citep{Gibb_ApJSS_2004, Oberg_2011, vanDishoeck2013}, and it has been the subject of extensive laboratory studies, mostly in the mid- and far-infrared spectral range by Fourier transform infrared spectroscopy (FTIR) \citep[][among others]{Bertie_JCP_1967, Bertie_JCP_1969, Hagen_CP_1981, JCP.78.11.6399.1983, Warren_AO_1984, Hudgins1993, Clapp_JPC_1995, Curtis_AO_2005, AJ.701.2.1347.2009, APJ.401.353.1992, AA.103.45.1994, Toon_JGR_1994, AA.565.A108.2014, Bouilloud_MNRAS_2015, Rocha_AA_2022, Mifsud2022, Hudson2025}. Fewer studies are available in the UV-vis regime \citep{Kofman_ApJ_2019, He_ApJ_2022}. Some studies have focused on the analysis of the structural features and the porosity of ices in relation to deposition conditions \citep{Westley1998, Stevenson1999, Dohnalek2003, Mason2006, Escribano_AA_sep2025} as well as on modelling using computational methods \citep{Moberg_JPCL_2017, Escribano_AA_2025} and the development of algorithms for the determination of the optical constants of ice samples relevant for astrophysical environments \citep{Gerakines2020ModifiedAlgorithm, Rocha_AA_2024}.

Water ice spectroscopic properties in the THz regime are less constrained, with the most reliable THz spectra recently reported for the Ih ice by \cite{Tao_JCP_2024}. A probable reason for the lack of spectroscopic data in the THz region is the significantly lower absorption band strength of water ice, by four to five orders of magnitude, compared to the IR region. This characteristic requires more demanding experimental conditions, particularly in terms of the ice sample thicknesses necessary to account for the dynamic range and sensitivity of the spectroscopic systems \citep{Curtis_AO_2005}. Therefore, an extension of laboratory characterisation of water ice is needed, especially in the THz region.

Following the methodology presented in our previous investigation of \ce{CO}, \ce{CO2}, and \ce{N2} ices \citep{AA.629.A112.2019, AA.667.A49.2022, AA.701.A287.2025, AA_N2_2025}, we took advantage of THz pulsed spectroscopy (TPS) to measure the THz response of water ice directly without the need for a Kramers–Kronig transform. The Fourier transform of the THz waveform transmitted through the sample gives both the spectral amplitude and phase, allowing for retrieval of its complex refractive index. Moreover, the TPS spectral phase allows for calibration of the spectral phase of FTIR data, eliminating the uncertainty of the Kramers-Kronig transform for the broadband characterisation of ices \citep{AA.667.A49.2022}.

Water ice characteristics, such as density and physical state, are a result of different growth methods and conditions, including pressure, thermal history, and substrate \citep{Stevenson1999, Kimmel2001, Salzmann_JCP_2019, Loerting_CC_2020}. In the context of the cold ISM, amorphous solid water (ASW) seems to be a more likely form of water in icy dust grains of molecular clouds \citep{AA.79.256.1979, Boogert_ApJ_2008, ARAA.52.1.541.2015, Perotti_AA_2020}, while crystalline phases are also possible \citep{Dartois_AA_2002, Terada_ApJ_2012}. 

\begin{figure}[!t]
    \centering
    \includegraphics[width=1.0\columnwidth]{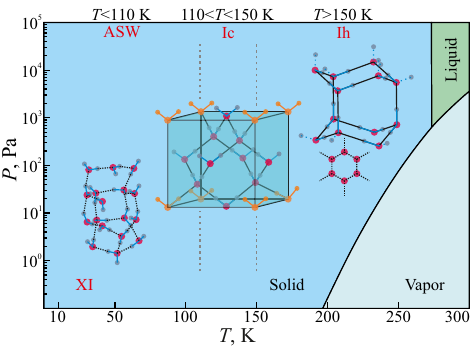}
    \caption{Fragment of
    the phase diagram of water ice
    in the temperature and pressure ranges
    relevant for the cold interstellar environments.}
    \label{FIG:PhaseDiagram}
\end{figure}

In Fig.~\ref{FIG:PhaseDiagram}, a phase diagram of water is shown. Although $19$ crystalline and four amorphous phases can generally be obtained in a laboratory within the temperature and pressure ranges relevant to cold interstellar environments, a much smaller diversity is expected. The main ice forms that are believed to be present in astrophysical environments are the following:

\begin{itemize}

\item In the Ih ice oxygen atoms form hexagonal symmetry, with close-to-tetrahedral bonding angles. This phase can be obtained by freezing liquid water \citep{Tao_JCP_2024} or through vapour deposition at temperatures equal to or above $150$~K \citep{Curtis_AO_2005, Rocha_AA_2024, Escribano_AA_2025}.

\item The Ic ice is a metastable crystalline phase with the cubic lattice \citep{NM.19.586.2020}, which can be formed by gas-phase deposition at temperatures ranging from $110$ to $150$~K \citep{Curtis_AO_2005, Rocha_AA_2024, Escribano_AA_2025}. It is commonly observed at temperatures above $120$~K as a mixture of Ic and ASW ices \citep{AJ.701.2.1347.2009}.

\item There are two compact phases of ASW at temperatures below $120$~K \citep{Carmack_ApJ_2023}: the high-density ASW is obtained at temperatures below $40$~K, whereas deposition between $40$ and $120$~K yields the low-density ASW form \citep{Narten_JCP_1976, Jenniskens1994}. In addition, ices obtained with the background deposition technique become porous at lower temperatures. Therefore, the two forms, generally referred to as porous ASW (p-ASW) and compact ASW (c-ASW), are also distinguished \citep{Stevenson1999, Dohnalek2003, Collings2003, Eklund2026}.

\end{itemize}

The THz--IR response of \ce{H2O} ice depends on phase, density, and porosity properties, which are functions of the deposition conditions. For ASW, this dependence is especially pronounced \citep{Dohnalek2003, PNAS.104.47.18387.2007, AJ.758.1.17.2012, PCCP.15.10.3630.2013, SR.3.3005.2013}, with a faster deposition resulting in more-porous ices \citep{Escribano_AA_2025}.
In this work, we analyse the broadband optical constants of Ih, Ic, and ASW \ce{H2O} ices; measure their THz optical properties with our custom-designed TPS experimental setup; and compare them to available literature data. We merge their broadband optical properties in the frequency range of $\nu = 0.3$--$120$~THz ($\lambda = 1$~mm--$2.5$~$\mu$m, respectively), and we model the observed vibrational absorption bands by a multiple-Lorentz model. We examine the observed discrepancies between these data and connect them to the porosity of water ice.

\section{Methods}
\label{SEC:Meth}

\subsection{The experimental setup}
\label{SEC:Cryo}

The THz response of water ice samples is studied experimentally at the CASICE laboratory at the Centre for Astrochemical Studies, Max Planck Institute for Extraterrestrial Physics (Garching, Germany). The experimental setup is based on an Advanced Research Systems cryocooler and a BATOP TDS-$1008$ TPS spectrometer and has been customised to our design.
The method for quantifying the THz optical properties of ices relying on the TPS signals was developed by \citet{AA.629.A112.2019} and has already been applied in studies of different ices \citep{AA.667.A49.2022, AA.701.A287.2025, AA_N2_2025}.

The $15$-cm diameter vacuum chamber is equipped with a high-power closed-cycle cryocooler that cools the sample holder down to a temperature as low as $5$~K. The base pressure of the system is maintained to $\simeq 10^{-7}$~mbar (when cold) by a pumping station made of a turbomolecular pump (84 ls$^{-1}$ nitrogen pumping speed) and a backing rotary pump (5 m$^3$ h$^{-1}$ pumping speed).
The substrate placed in the middle of the vacuum chamber and the optical windows of the chamber are made of high-resistivity float-zone silicon (HRFZ-Si), with a high refractive index ($n_\mathrm{Si} \approx 3.4$) and negligible dispersion and absorption in the THz--IR range. The use of the HRFZ-Si substrate is crucial for measurements at different temperatures, as its THz-IR response is nearly temperature independent.

The TPS system utilises a pair of photoconductive antennas pumped and probed by a femtosecond fibre laser (TOPTICA FemtoFErb $780$) to emit and detect broadband THz pulses. It exhibits good transmittance in the spectral range of $\nu = 0.05$--$3.5$~THz and a spectral resolution as high as $\Delta \nu = 0.03$~THz.
The sample compartment was customised to accommodate the cryocooler. The optical path within it is purged with cold nitrogen gas to suppress the impact of atmospheric water on the measured data along the THz beam path.

\subsection{The experimental protocol}
\label{SEC:IcePrep}

Following the experimental procedure by \citet{AA.629.A112.2019}, the ice layers were formed by gas expansion into the vacuum chamber through a $6$-mm diameter stainless steel pipe spaced at a distance of $\simeq 7$~cm away from the substrate, followed by the condensation of gas phase onto the HRFZ-Si substrate. This ensured a uniform background ice deposition on both sides of the substrate.
The water sample was degassed via at least three freeze-pump-thaw cycles before being expanded into the cryostat's vacuum chamber.
Water ice samples with a thickness of a few hundred microns are required for TPS characterisation. For this reason, the chosen deposition rate had to be fast enough to grow thick ices in a reasonable amount of time, resulting in a chosen deposition rate of several $\mu$m per minute.

The pressure inside the vacuum chamber did not exceed $\approx 5 \times 10^{-7}$~mbar during ice deposition, which  was performed in steps. Each deposition time step was set to $30$~min or $60$~min. During this time, the temperature of the HRFZ-Si substrate increased by a maximum of $\approx 2$~K due to the release of heat from the gas load condensation and ice growth onto the substrate.
After each deposition step, the experimental system was allowed enough time to return to its base temperature and pressure conditions before the TPS spectrum acquisition. Ice growth was monitored by measuring the time delays between the THz pulses from the current and previous deposition steps.

The deposition of ices was carried out at the substrate base temperatures of $T = 150$, $120$, and $8$~K, which most likely resulted
in the expected Ih, Ic, and ASW ices, respectively \citep{Curtis_AO_2005, Rocha_AA_2024}.
For the Ih, Ic, and ASW ices, the total deposition time was $\simeq 300$, $270$, and $360$~min, respectively, leading to a maximal ice thicknesses of $\simeq 859$ and $621$~$\mu$m, $\simeq 676$ and $485$~$\mu$m, and $\simeq 1082$ and $886$~$\mu$m on each side of the substrate. A precise estimate of the error associated with the thickness calculation is difficult, but we estimate it to be of the order of a few microns.

\subsection{THz optical properties
and porosity of ices}

In accordance with previous papers within our series \citep{AA.629.A112.2019, AA.667.A49.2022, AA.701.A287.2025, AA_N2_2025}, the THz--IR response of ices is described by the refractive index $n$ and the absorption coefficient $\alpha$. These are related to the complex refractive index $\tilde{n} = n' - i n''$ and the complex dielectric permittivity
$\tilde{\varepsilon} = \varepsilon' - i \varepsilon''$
via
\begin{equation}
    \tilde{n}= \sqrt{ \tilde{\varepsilon} } \equiv n - i \frac{ c_\mathrm{0} } { 2 \pi \nu } \alpha,
    \end{equation}
where $n \equiv n'$, $\nu$ stands for the frequency and $c_\mathrm{0}$ is the speed of light in vacuum.

As discussed by \cite{AA.701.A287.2025}, to define the relation between the dielectric response of compact and porous analytes, the Bruggeman effective medium model \citep{Bruggeman1935} was applied. Once an analysed medium is a compact material filled with empty pores, this model takes the form
\begin{equation}
    \left( 1 - P \right) \frac{ \tilde{\varepsilon}_\mathrm{bulk} -\tilde{\varepsilon} }{\tilde{\varepsilon}_\mathrm{bulk} + 2 \tilde{\varepsilon} } + P \frac{ \varepsilon_\mathrm{pore} - \tilde{\varepsilon} }{\varepsilon_\mathrm{pore} + 2 \tilde{\varepsilon} }
    = 0,
    \label{EQ:Bruggeman}    
\end{equation}
where $\tilde{\varepsilon}$ is the effective complex dielectric permittivity of a porous medium. Further, $\varepsilon_\mathrm{pore} = 1$ and $P$ are the dielectric permittivity and volume fractions of pores (free space), respectively, while $\tilde{\varepsilon}_\mathrm{bulk}$ and $\left( 1 - P \right)$ are the dielectric permittivity and volume fractions of the compact material.
The porosity value, $P$, is obtained by solving this equation using experimental data for $\tilde{\varepsilon}$ and literature values, $\tilde{\varepsilon}_\mathrm{bulk}$, for compact ice. 
The accuracy of the derived value, which can be robustly estimated as $\pm$1.5\% in the present work, depends on the accuracy of the measurements of the optical properties for bulk and porous ices.
To retrieve the scattering coefficient ($\mu$) of ices and estimate the underlying effective radius ($R_\mathrm{eff}$) of pores from the measured optical constants, we used the method developed by \cite{AA.701.A287.2025}, which combines the Lorentz-Mie scattering theory and the radiative transfer theory.

\subsection{Multiple-Lorentz model for the complex dielectric permittivity of ices}

In a similar manner to our previous papers \citep{AA.629.A112.2019, AA.667.A49.2022, AA.701.A287.2025, AA_N2_2025}, the absorption bands of ices are described by the classical multiple-Lorentz model—a superposition of the Lorentz terms
\begin{equation}
    \widetilde{\varepsilon}\left( \nu \right)
    = \varepsilon_\infty
    + \sum_{j=1}^{N_\mathrm{L}} \frac{ \Delta\varepsilon_j\nu_{\mathrm{L},j}^{2} }
    { \nu^{2}_{\mathrm{L},j} - \nu^{2}  + i \nu \gamma_{\mathrm{L},j} }\:,
    \label{EQ:LorentzModel}
\end{equation}
where $\Delta\varepsilon_j$, $\nu_{\mathrm{L},j}$, and $\gamma_{\mathrm{L},j}$ are the real amplitude, resonant frequency, and damping constant of the $j^\mathrm{th}$ oscillator and $\varepsilon_\infty$ is the real constant dielectric permittivity at higher frequencies.
In fact, $\nu_{\mathrm{L},j}$ and $\gamma_{\mathrm{L},j}$ define the position and width of the $j^\mathrm{th}$ absorption peak, while $\Delta \varepsilon_j$ regulates its contribution to the complex dielectric permittivity.

Since the absorption bands of water ice are reportedly non-Lorentzian and comprise a set of vibrational modes (for instance, it follows from the ab initio simulations by \cite{Moberg_JPCL_2017} and \cite{Escribano_AA_sep2025}), several Lorentz kernels are used to describe each band.
In contrast to the common series of Gaussian bands often used to decompose the FTIR spectra \citep{ARAA.52.1.541.2015}, the described model is physically rigorous.
Indeed, it satisfies the Kramers-Kronig relations and the sum rule \citep{PR.161.1.143.1967} originating from the causality principle and the charge conservation law, respectively, and governing the electrodynamic response of any physical system.
Moreover, the model is close to both the semi-analytical convolution models used to calculate the dielectric spectra relying on the ab initio simulations (where the Lorentz kernel is convolved with the discrete set of vibrational modes \citep{Moberg_JPCL_2017, Escribano_AA_sep2025}) and the fully analytical models, such as that by
\cite{JNCS.203.1.1996} (with the continuous density of states defined by Gaussian statistics).

\section{Results}
\label{SEC:Results}

\subsection{Literature data on the THz--IR optical properties of water ice}

\begin{figure*}[!t]
    \centering
    \includegraphics[width=2.0\columnwidth]{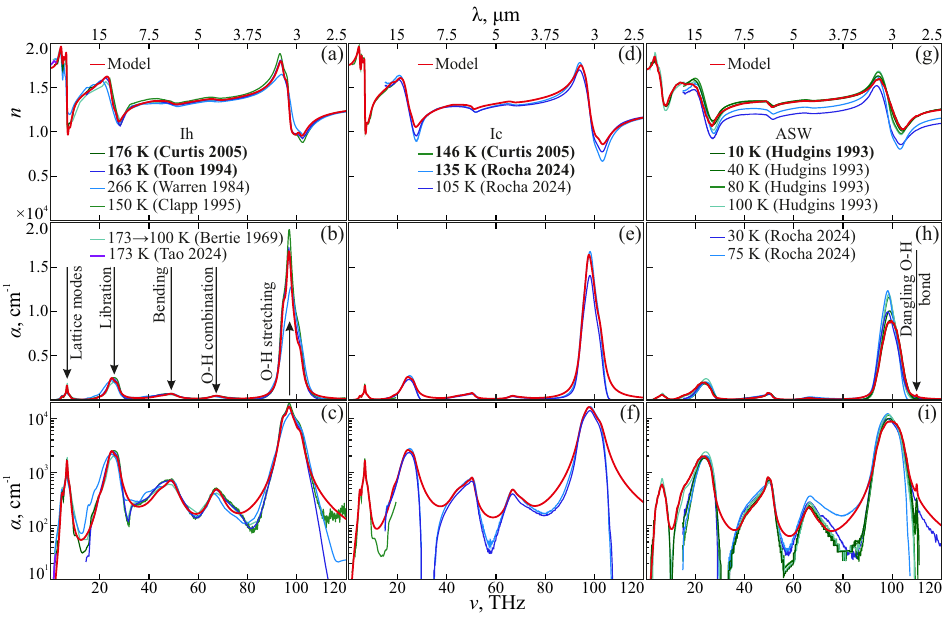}
    \caption{Literature data on the broadband (THz--IR) optical properties of \ce{H2O} ices as compared to the inferred model
    (Eq.~\eqref{EQ:LorentzModel} and Appendix~\ref{SEC:Appendix},
    Tabs.~\ref{TAB_Ih_Lorentz}--\ref{TAB_ASW_Lorentz}).
    (a)--(c): Literature data on the refractive index $n$ and absorption coefficient $\alpha$
    (by field, in both the linear and logarithmic scale), respectively, for the Ih ice overlapped with the model.
    The legend indicates the ice deposition (or warming-up) temperature and the corresponding reference. The bold text denotes the densest ice samples. The vertical arrows indicate the absorption bands assigned to the different vibrational modes.
    (d)--(f) and (g)--(i): Similar datasets for the Ic and ASW ices, respectively.}
    \label{FIG:LitModel}
\end{figure*}

In Fig.~\ref{FIG:LitModel}, we plot the literature data on the THz--IR optical properties of \ce{H2O} ice. Panels (a)--(c) illustrate the data relative to the Ih phase \citep{Warren_AO_1984, Toon_JGR_1994, Clapp_JPC_1995, Curtis_AO_2005, Tao_JCP_2024, Bertie_JCP_1969}, panels (d)--(f) show the Ic phase \citep{Curtis_AO_2005, Rocha_AA_2024}, and panels (g)--(i) are about the data of ASW samples \citep{Hudgins1993, Rocha_AA_2024}. 

For each set of data, the spectral refractive index, $n$, and the absorption coefficient, $\alpha$, are shown in linear and logarithmic scales.
Ice samples with the highest values of optical constants are believed to have the highest density values. We highlight them in the figure legend with bold text for the Ih \citep{Curtis_AO_2005, Toon_JGR_1994}, Ic \citep{Curtis_AO_2005, Rocha_AA_2024}, and ASW \citep{Hudgins1993} ices.
For all water ice samples, five absorption bands are visible, and we assigned them to the lattice (around $\nu \simeq 6$~THz), libration ($22$~THz), bending ($44$~THz), \ce{O-H} combination ($67$~THz), and \ce{O-H} stretching ($97$~THz) modes \citep{Curtis_AO_2005, Rocha_AA_2024}. For ASW only, an additional absorption can be observed at $\simeq 110$~THz, which we attributed to the free \ce{O-H} stretching mode, also known as the \ce{O-H} dangling mode \citep{Hsieh_PNAS_2013, Satorre_PCCP_2021, Nagasawa2021, Sudera_2020}. The \ce{O-H} dangling mode enables distinction between compact and porous amorphous water ice in both astrophysical observations and laboratory data \citep{Palumbo_2005, Noble_NatAst_2024, Nagasawa2021}. The absorption coefficient for the entire set of data decreases overall with frequency, reaching a value close to $n \simeq 1.8$ at THz frequencies.

To model the complex dielectric permittivity of the ice samples, we decomposed the data into Lorentz kernels (Eq.~\eqref{EQ:LorentzModel}). The results are overlapped with the literature curves in Fig.~\ref{FIG:LitModel}, while their parameters are given in Appendix~\ref{SEC:Appendix} (Tabs.~\ref{TAB_Ih_Lorentz}-- \ref{TAB_ASW_Lorentz}).
From panels~(a),(d), and (g), we noticed that the developed model overall reproduces the broadband refractive indices. It describes well the high values of absorption near the absorption peaks in the linear scale plots ((b),(e,) and~(h)), while it fails to describe regions of low absorption between the peaks (baseline) visible in the logarithmic plots ((c),(f), and (i)).
The model exhibits a broadening of the absorption bands from Ih to ASW ices (i.e. an increase in the number of Lorentz terms required to describe each band), which is not visible in the experimental data.

The difficulty the simple Lorentz model has in describing the low absorption values between the peaks as well as the asymmetric and broadened peaks can be attributed to different factors.
In this model, the damping is constant, leading to peak symmetry. However, due to anharmonicity, the actual bands may differ from this ideal case \citep{PR.155.3.882.1967}.
When dealing with non-Lorentzian peaks, one can resort to more complex models, such as the coupled oscillator \citep{PR.135.6A.1964} or four-parameter factorised \citep{PR.174.3.791.1968, PRB.22.11.5501.1980, NC.7.10193.2016} models, but their applicability is limited by a large number of model parameters, which do not always reflect the physics and chemistry of an analyte.

Another reason for the less accurate match between the data and the model could be attributed to the baseline subtraction procedure in the experimental spectra, which is often applied in FTIR spectroscopy but is usually not described in detail in articles.
This procedure may filter out important information about interactions between electromagnetic waves and a sample, such as scattering effects, interference patterns, and non-local low-intense absorption bands (for example, the Boson peaks). Thus, the baseline subtraction can affect the useful information in the measured data by distorting the shape and absolute values of the absorption bands, as evident in panels~(c),(a), and~(i), which plot data by \cite{Rocha_AA_2024}.

\subsection{Measured THz optical properties and the porosity of water ice}

\begin{figure*}[!t]
    \centering
    \includegraphics[width=2.0\columnwidth]{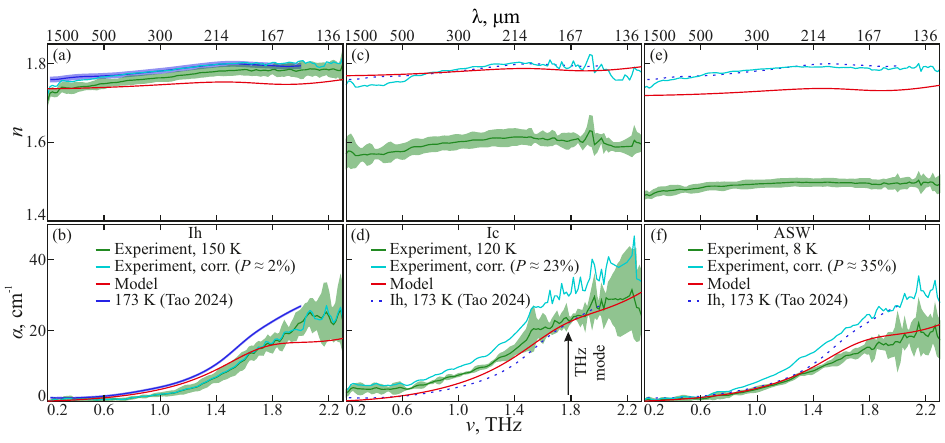}
    \caption{Experimental (this work) and literature data on the THz optical properties of \ce{H2O} ices as compared to the inferred model (Eq.~\eqref{EQ:LorentzModel} and Appendix~\ref{SEC:Appendix}, Tabs.~\ref{TAB_Ih_Lorentz}--\ref{TAB_ASW_Lorentz}).
    (a),(b): Measured refractive index ($n$) and absorption coefficient ($\alpha$; by field), respectively, for the Ih ice overlapped with the data by \cite{Tao_JCP_2024} and the model.
    The legend indicates the ice deposition temperature and the corresponding reference.
    The blue curves show the experimental data corrected by the Bruggeman model (Eq.~\eqref{EQ:Bruggeman}) to exclude the $P = 2\%$ ice porosity (in comparison to data by \cite{Tao_JCP_2024}) as detailed by \cite{AA.701.A287.2025}.
    (c),(d) and (e),(f): Similar datasets for the Ic and ASW ices, respectively.
    In the absence of relevant data in the literature, the data for Ih and ASW ices are also compared with those (dashed curves) for Ih ice by \cite{Tao_JCP_2024}, with the resultant porosity values constituting $P = 23\%$ and $35\%$ for our Ih and ASW samples, respectively.}
    \label{FIG:IceResponse}
\end{figure*}

In Fig.~\ref{FIG:IceResponse}, we show the mean values of the measured refractive index and absorption coefficient of water ice in the $0.2$--$2.3$~THz spectroscopic region (for which few literature data are available) together with the $\pm 1.5 \sigma$ error bars (where $\sigma$ is the standard deviation).
Panels~(a) and~(b) illustrate the experimental curves for our vapour-deposited Ih ice samples, the inferred model, and the data for the compact Ih ice obtained by \cite{Tao_JCP_2024} on samples of frozen water. The THz optical properties of our vapour-deposited Ih ice are quite similar to those obtained by freezing water. No narrow spectral features are visible, but a change in the slope of the baseline shape around $1.6$~THz has been interpreted by \cite{Tao_JCP_2024} as a broad feature of low spectral intensity.
Following the same interpretation, in our model, we describe the broad THz band around $1.8$~THz by an additional Lorentz term with $\Delta \varepsilon_\mathrm{L} = 0.049$, $\nu_\mathrm{L}=1.78$~THz, and $\gamma_\mathrm{L}=1.0$~THz (Eq.~\eqref{EQ:LorentzModel} and Appendix~\ref{SEC:Appendix}).

Assuming the frozen-water Ih ice in the experiment by \cite{Tao_JCP_2024} is compact, we applied Eq.~\eqref{EQ:Bruggeman} to our vapour-deposited Ih ice as described by \cite{AA.701.A287.2025} and estimated its porosity to be as low as $P = 2$\%.
We then used this porosity value and Eq.~\eqref{EQ:Bruggeman} to correct our experimental data and eliminate the contribution of pores to the THz optical properties of the sample. The resultant corrected curves are shown with dashed blue lines in Fig.~\ref{FIG:IceResponse} (a) and~(b), and they are very close to the initial curves.
The measured THz optical properties before and after correction as well as those predicted by our model show good agreement. Our model slightly underestimates the THz refractive index, which can be explained by its reliance on IR literature data for vapour-deposited Ih ice, which is treated as compact but may still retain residual porosity.

When comparing the corrected experimental curve (dashed blue line) with our model (solid red line) at increasing frequencies, we observed that the experimental data exhibit excessive extinction. We attribute this effect to THz-wave scattering from pores and analysed it using the methods developed by \cite{AA.701.A287.2025}.
In short, we subtracted the model absorption from the corrected experimental values and fit the resultant curve using the scattering coefficient model defined in the Rayleigh limit as $\mu_\mathrm{s} = A \nu^{4},$ where $A$ is a function of the known volume fraction $P$ and effective radius $R_\mathrm{eff}$ of pores, 
while $\varepsilon_\mathrm{bulk}$ is the dielectric constant of a bulk host medium.
This yielded a reasonably effective pore radius of $R_\mathrm{eff} = 13.1$~$\mu$m for Ih ice, which underlies the observed scattering effects.

In Fig.~\ref{FIG:IceResponse} panels~(c),(d) and~(e),(f), the data for the vapour-deposited Ic and ASW ices are shown. In the absence of THz literature data on the compact Ic and ASW ices, we compared our vapour-deposited samples with the Ih ice reported by \cite{Tao_JCP_2024} and a model that includes a broad THz feature around $1.8$~THz.
While the shapes of the curves for the different ice samples are quite similar, the values of the THz optical constants measured for the Ic and ASW ices are much lower than those for compact Ih. We attribute this effect to the porosity of our vapour-deposited Ic and ASW ice samples, and we quantified it in a manner similar to that of Ih ice above.

For the Ic and ASW ices, the porosity appears to be as high as $P \approx 23$\% and $35$\%, respectively.
For both ices, the inferred model predicts the same shape of the THz spectra, while their values are close to the compact Ih ice. Therefore, we corrected the experimental curves to simulate a compact ice sample and then compared the excess absorption observed in the corrected curves (dashed blue line) with the model (solid red line).
This analysis resulted in the effective pore radii being as high as $R_\mathrm{eff} = 13.8$ and $11.1$~$\mu$m for the vapour-deposited Ic and ASW ices, respectively. 
The error in pore size determination depends on many factors, 
including the accuracy of the experimental scattering coefficient fitting procedure, the
accuracy in porosity determination, and the accuracy of known dielectric properties of compact ice. 
In this study, uncertainties in determining the effective pore radii are expected to be as small as $\pm0.5$~$\mu$m.
These $R_\mathrm{eff}$ values are quite reasonable for extremely thick (a few hundred $\mu$m) ice samples, which are necessary for measurements in the THz range.
For the Ic and ASW ices, almost all discrepancies between the measured data and the model are explained by the porosity and scattering properties of ices.
In turn, the non-zero absorption of Ic ice at lower frequencies cannot be attributed to scattering by pores.
It might be caused by imperfections in the ice sample geometry as compared to the plane-parallel one, the presence of a meniscus, or wedges.
In fact, they cannot be accounted for during spectroscopic data processing.

\subsection{Integrity of the modelled THz--IR optical properties of water ice}

The integrity of the modelled THz--IR optical properties of water ices can be analysed in the context of the sum rule \citep{PR.161.1.143.1967}, assuming that the considered spectral range is broad enough to span all optically active vibrational modes of an analyte.
According to the sum rule, the integral over the frequency-dependent dynamic conductivity $\sigma = 2 \pi\nu \varepsilon_0 \varepsilon''$ is directly proportional to the number of dipoles or charges underlying the spectral resonances of an analyte.

Taking into account only the dynamic conductivity associated with the vibrational modes, this integral is clearly defined by the strengths of the Lorentz oscillators $\Delta \varepsilon_j \nu_{\mathrm{L},j}$, and it should be constant for all forms of water ice:
\begin{equation}
    \int_{0}^\infty \sigma d \nu
    \propto \sum_j \Delta \varepsilon_j \nu_{\mathrm{L},j}^{2}
    = \mathrm{const}.
    \label{EQ:SumRule}
\end{equation}
In this way, the sum rule demands some equal integral absorptivity of ices.
When Eq.~\eqref{EQ:SumRule} is applied to the inferred models (Eq.~\eqref{EQ:LorentzModel} and Appendix~\ref{SEC:Appendix}, Tabs.~\ref{TAB_Ih_Lorentz}--\ref{TAB_ASW_Lorentz}), the resultant total strength of oscillators varies by $\simeq30\%$ for the considered Ih, Ic, and ASW ices.
This difference can be explained by the aforementioned baseline subtraction that non-physically reduces the dynamic conductivity $\sigma$.

\subsection{Opacities of dust grains covered with water ice}

\begin{figure}[!t]
    \centering
    \includegraphics[width=1.0\columnwidth]{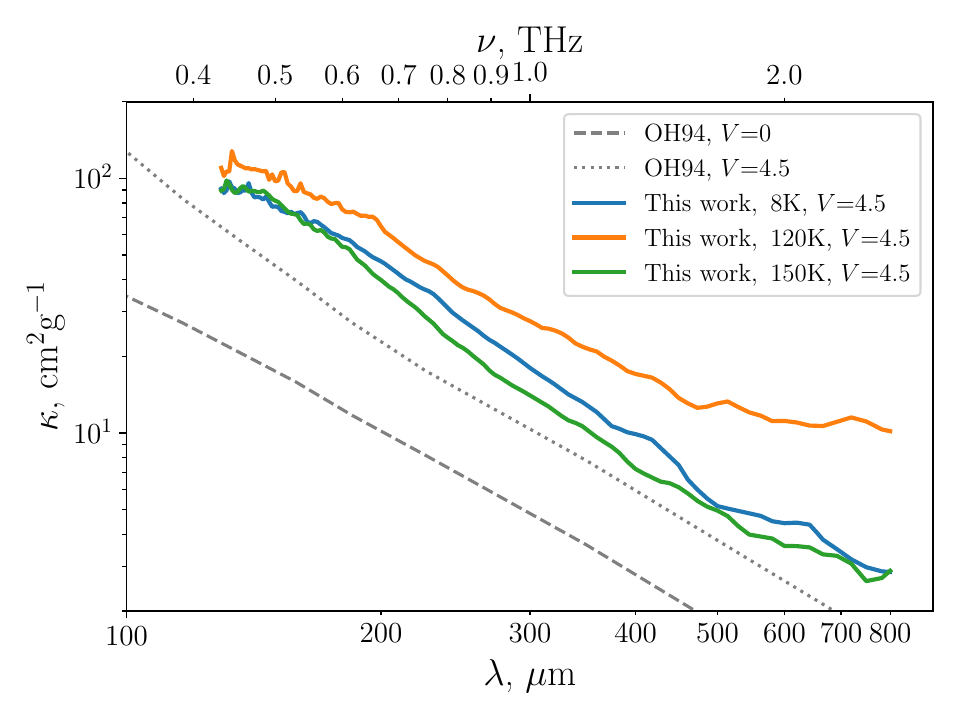}
    \caption{Calculated and reference opacities of astrophysical dust, plotted as a function of the wavelength. Dashed and dotted grey lines labelled with OH94 refer to bare grains and dust grains with icy mantles employing the optical properties described in \citet{1994AA.291.943O}, respectively. Opacities for grains with pure \ce{H2O} icy mantles at different temperatures, computed for optical constants of the present work at different temperatures, are depicted by solid lines.}
    \label{FIG:kappa}
\end{figure}

The optical properties of grains play a key role in several astrophysical environments, not only for computing the opacity to include in the models (e.g. \citealt{Grassi2017,Ballering2021,Arabhavi2022,Dartois2022}) but also for determining their observability and their actual composition \citep{Boogert2015} and retrieval methods \citep{Rocha2021,Megias2025}.
Since ice-coated dust grains are expected to be relatively abundant in cold, dense, and UV-shielded environments—from dense molecular clouds and prestellar cores to the cold outer and midplane regions of protoplanetary disks— in Figure~\ref{FIG:kappa} we report\footnote{The data and the plotting routine are publicly available at \url{https://github.com/tgrassi/qabs_compute}; script \texttt{test\_07.py}, commit \texttt{200cf57}.} the opacity $\kappa$ computed from the optical constant of ASW, Ih, and Ic H$_2$O ices measured in this work at different temperatures, namely 8, 120, and 150\,K with a volume ratio\footnote{Given a spherical dust grain of radius $a$ covered with an additional layer of ice $\Delta a$, the volume ratio is $V=(1+\Delta a / a)^3-1.$} of ice of $V=4.5$, which is a reasonable assumption for the astrochemical ices. 

For comparison, we plot the opacities computed with the optical properties from \citet{1994AA.291.943O} for bare grains (i.e. zero volume ratio; dashed grey line) and for a coating of H$_2$O:CH$_3$OH:CO:NH$_3$=100:10:1:1 (dotted grey line), which is the mixture at 10\,K from \citet{Hudgins1993} with the addition of spherical amorphous carbon inclusions from \citet{1993A&A...279..577P} mixed with the Bruggeman effective medium approximation \citep{Bruggeman1935}. Details of the calculation employing Mie scattering theory \citep{1983asls.book.....B} assuming an MRN dust grain size distribution \citep{1977ApJ...217..425M} can be found in \citet{AA.629.A112.2019}.

In addition to the wavelength-dependent opacity data, we provide a power-law fit, $\kappa = \kappa_0 \left(\lambda / 1\, \mu{\rm m}\right)^{-\beta}$, traditionally employed to describe the opacity at $\lambda \gtrsim 100\,\mu$m \citep{1994AA.291.943O,Ormel2011,2014prpl.conf..339T}. The values of the parameters of the power-law fit $\kappa$ at different ice deposition temperatures are listed in Table~\ref{beta}. In protoplanetary disks, $\beta<1$ is expected \citep{Draine2006,Birnstiel2012}, while larger values up to $\beta\approx2$ are found in objects such as molecular clouds and cores \citep{Juvela2018}. However, several exceptions suggest that a characterisation of the dust's optical properties is crucial for understanding them \citep{Nozari2025}.

\begin{table}[!t]
\caption{Parameters of the power-law fit $\kappa$, defined in the text, for the different ice deposition temperatures.}
\renewcommand{\arraystretch}{1.5}
 \begin{tabularx}{\linewidth}{lll}
 $T$ (K) & $\kappa_0$ & $\beta$\\
\hline
8   & 3.04(6) & 2.10\\
120 & 2.81(5) & 1.60\\
150 & 5.06(6) & 2.22\\
\hline
 \end{tabularx}
 \tablefoot{Note that $a(b)=a\times10^b$.}
 \label{beta}
\end{table}

\section{Discussion and conclusions}

In the present work, we have determined the extended THz--IR broadband optical properties of the Ih, Ic, and ASW ices. Data obtained from the literature and measured experimentally by TPS were merged in the broad frequency range of $\nu = 0.3$--$120$~THz (the wavelength range of $\lambda = 1$~mm--$2.5$~$\mu$m). This wide spectral region accommodates most of the optically active vibrational modes of ices. The absorption bands were modelled in a unified manner by the multiple-Lorentz model of complex dielectric permittivity (see Eq.~\eqref{EQ:LorentzModel} and Appendix~\ref{SEC:Appendix}), where each band was described by a set of Lorentz terms to account for the anharmonicity effects.

The main challenges encountered in modelling the FTIR data are related to the correction of the background absorption and the inability of the classical Lorentz model to describe wings of the absorption peaks, while in the THz region, the main issues in the optical constant derivation are the porosity and scattering properties of our thick ice samples. In particular, the porosity of the water ice was found to increase in the $P = 2\%$ -- $35$\% range, with the substrate temperature varying from $T = 150$ to $8$~K. This might introduce uncertainties into astrophysical predictions of the absorptivity and emissivity of icy dust grains. To minimise this effect, we applied a correction to the measured data that accounts for the different porosity levels of the ice samples. This should both exclude any non-physical data processing and provide error bars and detectable absorption levels.

For all states of water ice, the retrieved broadband optical properties are very close. In the THz region, the largest discrepancy is observed for the Ic ice data, which could be explained by some variations in the phase state and the porosity of the sample. For an ice film deposited from the gas phase onto a cold substrate, one can expect a temperature gradient across the film thickness during deposition. In fact, this effect is usually observed for the Ic ice, which is commonly mixed with ASW \citep{AJ.701.2.1347.2009}.

The THz spectra do not provide useful information for the discrimination of the distinct forms of water ice (Fig.~\ref{FIG:IceResponse}).
For this, the IR range is preferable, thanks to the dangling \ce{O-H} mode of ASW ($110$~THz) \citep{Palumbo_2005,  Hsieh_PNAS_2013, Sudera_2020, Satorre_PCCP_2021, Nagasawa2021, Noble_NatAst_2024}, as well as observed changes in the profile of the absorption bands with the deposition conditions for the lattice modes ($6$~THz), libration ($22$~THz), bending ($44$~THz), \ce{O-H} combination ($67$~THz), and \ce{O-H} stretching ($97$~THz) bands, (Fig.~\ref{FIG:LitModel}, Eq.~\eqref{EQ:LorentzModel}, and Appendix~\ref{SEC:Appendix}).
On the other hand, the TPS data provide accurate measurements of ice dielectric properties in the THz range, providing direct reconstruction of the complex dielectric function of ices without the use of the Kramers–Kronig relations.

In cold, dense regions, the opacities of dust grains are influenced by the spectral features of the ice mantles covering them, depending on their chemical composition. Accurate values of opacity are therefore necessary to interpret dust continuum observations, especially in star- and planet-forming regions and at the snowlines of protoplanetary disks.
Calculating the opacities of dust grains covered by \ce{H2O} ice mantles using our experimental data produces results that differ from those derived by extending IR data. This discrepancy cannot be fully attributed to the mixed composition of the ice used in \citet{1994AA.291.943O}, as the effect of the minor components on the optical constants of the ice mixture should be negligible. Therefore, our findings indicate that laboratory measurements provide a more accurate interpretation of dust continuum emission.

\begin{acknowledgements}
The authors appreciate the support of the Max Planck Society. This project has received funding from the European Union’s Horizon~$2020$ research and innovation program under the Marie Skłodowska-Curie grant agreement \#~$811312$ for the Project ''Astro-Chemical Origins'' (ACO).
F.K. acknowledges support from the European Union’s Horizon Europe research and innovation programme under the Marie Skłodowska-Curie Actions Postdoctoral Fellowship grant agreement No. \# 101153804 (ORCHID).
IR and THz data analysis by A.A.G. and M.K.M. were supported by the RSF~Project \#~$25$--$72$--$00139$.

\end{acknowledgements}

\bibliographystyle{aa} 
\bibliography{REFs}

\begin{thebibliography}{96}
\expandafter\ifx\csname natexlab\endcsname\relax\def\natexlab#1{#1}\fi

\bibitem[{{Andrews} {et~al.}(2018){Andrews}, {Huang}, {P{\'e}rez}, {Isella}, {Dullemond}, {Kurtovic}, {Guzm{\'a}n}, {Carpenter}, {Wilner}, {Zhang}, {Zhu}, {Birnstiel}, {Bai}, {Benisty}, {Hughes}, {{\"O}berg}, \& {Ricci}}]{Andrews2018}
{Andrews}, S.~M., {Huang}, J., {P{\'e}rez}, L.~M., {et~al.} 2018, \apjl, 869, L41

\bibitem[{Arabhavi {et~al.}(2022)Arabhavi, Woitke, Cazaux, Kamp, Rab, \& Thi}]{Arabhavi2022}
Arabhavi, A., Woitke, P., Cazaux, S., {et~al.} 2022, \aap, 666, A139

\bibitem[{{Ballering} {et~al.}(2021){Ballering}, {Cleeves}, \& {Anderson}}]{Ballering2021}
{Ballering}, N.~P., {Cleeves}, L.~I., \& {Anderson}, D.~E. 2021, \apj, 920, 115

\bibitem[{Barker \& Hopfield(1964)}]{PR.135.6A.1964}
Barker, A. \& Hopfield, J. 1964, Physical Review, 135, A1732

\bibitem[{Berreman \& Unterwald(1968)}]{PR.174.3.791.1968}
Berreman, D. \& Unterwald, F. 1968, Physical Review, 174, 791

\bibitem[{Bertie \& Whalley(1967)}]{Bertie_JCP_1967}
Bertie, J. \& Whalley, E. 1967, \jcp, 46, 1271

\bibitem[{Bertie {et~al.}(1969)Bertie, Labbé, \& Whalley}]{Bertie_JCP_1969}
Bertie, J.~E., Labbé, H.~J., \& Whalley, E. 1969, The Journal of Chemical Physics, 50, 4501

\bibitem[{{Birnstiel} {et~al.}(2012){Birnstiel}, {Klahr}, \& {Ercolano}}]{Birnstiel2012}
{Birnstiel}, T., {Klahr}, H., \& {Ercolano}, B. 2012, \aap, 539, A148

\bibitem[{{Bohlin} {et~al.}(1978){Bohlin}, {Savage}, \& {Drake}}]{Bohlin_1978}
{Bohlin}, R.~C., {Savage}, B.~D., \& {Drake}, J.~F. 1978, \apj, 224, 132

\bibitem[{{Bohren} \& {Huffman}(1983)}]{1983asls.book.....B}
{Bohren}, C.~F. \& {Huffman}, D.~R. 1983, {Absorption and scattering of light by small particles}

\bibitem[{Boogert {et~al.}(2008)Boogert, Pontoppidan, Knez, Lahuis, Kessler-Silacci, van Dishoeck, Blake, Augereau, Bisschop, Bottinelli, Brooke, Brown, Crapsi, Evans~II, Fraser, Geers, Huard, J{\o}rgensen, {\"O}berg, Allen, Harvey, Koerner, Mundy, Padgett, Sargent, \& Stapelfeldt}]{Boogert_ApJ_2008}
Boogert, A., Pontoppidan, K., Knez, C., {et~al.} 2008, \apj, 678, 985

\bibitem[{Boogert {et~al.}(2015)Boogert, Gerakines, \& Whittet}]{ARAA.52.1.541.2015}
Boogert, A.~A., Gerakines, P.~A., \& Whittet, D. 2015, \araa, 53, 541

\bibitem[{{Boogert} {et~al.}(2015){Boogert}, {Gerakines}, \& {Whittet}}]{Boogert2015}
{Boogert}, A.~C.~A., {Gerakines}, P.~A., \& {Whittet}, D. C.~B. 2015, \araa, 53, 541

\bibitem[{Bouilloud {et~al.}(2015)Bouilloud, Fray, Bénilan, Cottin, Gazeau, \& Jolly}]{Bouilloud_MNRAS_2015}
Bouilloud, M., Fray, N., Bénilan, Y., {et~al.} 2015, Monthly Notices of the Royal Astronomical Society, 451, 2145

\bibitem[{Bruggeman(1935)}]{Bruggeman1935}
Bruggeman, D. 1935, Annalen der Physik, 416, 636

\bibitem[{Carmack {et~al.}(2023)Carmack, Tribbett, \& Loeffler}]{Carmack_ApJ_2023}
Carmack, R., Tribbett, P., \& Loeffler, M. 2023, The Astrophysical Journal, 942, 1

\bibitem[{{Caselli} {et~al.}(2019){Caselli}, {Pineda}, {Zhao}, {Walmsley}, {Keto}, {Tafalla}, {Chac{\'o}n-Tanarro}, {Bourke}, {Friesen}, {Galli}, \& {Padovani}}]{Caselli2019}
{Caselli}, P., {Pineda}, J.~E., {Zhao}, B., {et~al.} 2019, \apj, 874, 89

\bibitem[{{Caselli} {et~al.}(1999){Caselli}, {Walmsley}, {Tafalla}, {Dore}, \& {Myers}}]{Caselli_1999}
{Caselli}, P., {Walmsley}, C.~M., {Tafalla}, M., {Dore}, L., \& {Myers}, P.~C. 1999, \apjl, 523, L165

\bibitem[{{Clapp} {et~al.}(1995){Clapp}, {Worsnop}, \& {Miller}}]{Clapp_JPC_1995}
{Clapp}, M.~L., {Worsnop}, D.~R., \& {Miller}, R.~E. 1995, Journal of Physical Chemistry, 99, 6317

\bibitem[{Curtis {et~al.}(2005)Curtis, Rajaram, Toon, \& Tolbert}]{Curtis_AO_2005}
Curtis, D., Rajaram, B., Toon, O., \& Tolbert, M. 2005, Applied Optics, 44, 4102

\bibitem[{Dartois {et~al.}(2002)Dartois, d'Hendecourt, Thi, Pontoppidan, \& van Dishoeck}]{Dartois_AA_2002}
Dartois, E., d'Hendecourt, L., Thi, W., Pontoppidan, K., \& van Dishoeck, E. 2002, \aap, 394, 1057

\bibitem[{{Dartois} {et~al.}(2022){Dartois}, {Noble}, {Ysard}, {Demyk}, \& {Chabot}}]{Dartois2022}
{Dartois}, E., {Noble}, J.~A., {Ysard}, N., {Demyk}, K., \& {Chabot}, M. 2022, \aap, 666, A153

\bibitem[{Dohn{\'{a}}lek {et~al.}(2003)Dohn{\'{a}}lek, Kimmel, Ayotte, Smith, \& Kay}]{Dohnalek2003}
Dohn{\'{a}}lek, Z., Kimmel, G., Ayotte, P., Smith, R., \& Kay, B. 2003, The Journal of Chemical Physics, 118, 364

\bibitem[{{Draine}(2006)}]{Draine2006}
{Draine}, B.~T. 2006, \apj, 636, 1114

\bibitem[{{Dullemond} {et~al.}(2018){Dullemond}, {Birnstiel}, {Huang}, {Kurtovic}, {Andrews}, {Guzm{\'a}n}, {P{\'e}rez}, {Isella}, {Zhu}, {Benisty}, {Wilner}, {Bai}, {Carpenter}, {Zhang}, \& {Ricci}}]{Dullemond_2018}
{Dullemond}, C.~P., {Birnstiel}, T., {Huang}, J., {et~al.} 2018, \apjl, 869, L46

\bibitem[{{Dutrey} {et~al.}(1998){Dutrey}, {Guilloteau}, {Prato}, {Simon}, {Duvert}, {Schuster}, \& {Menard}}]{Dutrey_1998}
{Dutrey}, A., {Guilloteau}, S., {Prato}, L., {et~al.} 1998, \aap, 338, L63

\bibitem[{Efimov(1996)}]{JNCS.203.1.1996}
Efimov, A. 1996, Journal of Non-Crystalline Solids, 203, 1

\bibitem[{Elton \& Fernandez-Serra(2016)}]{NC.7.10193.2016}
Elton, D. \& Fernandez-Serra, M. 2016, Nature Commnications, 7, 10193

\bibitem[{Escribano {et~al.}(2025{\natexlab{a}})Escribano, del Burgo~Olivares, Carrascosa, Cazaux, Satorre, \& {Mu{\~n}oz~Caro}}]{Escribano_AA_2025}
Escribano, B., del Burgo~Olivares, C., Carrascosa, H., {et~al.} 2025{\natexlab{a}}, \aap, 699, A79

\bibitem[{Escribano {et~al.}(2025{\natexlab{b}})Escribano, del Burgo~Olivares, Carrascosa, Mart{\'\i}n-Dom{\'e}nech, G{\'o}mez, \& Mu{\~n}oz~Caro}]{Escribano_AA_sep2025}
Escribano, B., del Burgo~Olivares, C., Carrascosa, H., {et~al.} 2025{\natexlab{b}}, \aap, 701, A146

\bibitem[{Gavdush {et~al.}(2025)Gavdush, Ivlev, Zaytsev, Ulitko, Dolganova, Garnov, Giuliano, \& Caselli}]{AA.701.A287.2025}
Gavdush, A., Ivlev, A., Zaytsev, K., {et~al.} 2025, \aap, 701, A287

\bibitem[{Gavdush {et~al.}(2022)Gavdush, Kruczkiewicz, Giuliano, M{\"u}ller, Komandin, Grassi, Theul{\'e}, Zaytsev, Ivlev, \& Caselli}]{AA.667.A49.2022}
Gavdush, A., Kruczkiewicz, F., Giuliano, B., {et~al.} 2022, \aap, 667, A49

\bibitem[{Gerakines \& Hudson(2020)}]{Gerakines2020ModifiedAlgorithm}
Gerakines, P. \& Hudson, R. 2020, \apj, 901, 52

\bibitem[{Gibb {et~al.}(2004)Gibb, Whittet, Boogert, \& Tielens}]{Gibb_ApJSS_2004}
Gibb, E., Whittet, D., Boogert, A., \& Tielens, A. 2004, The Astrophysical Journal Supplement Series, 151, 35

\bibitem[{Giuliano {et~al.}(2014)Giuliano, Escribano, Martin-Domenech, Dartois, \& Mu\~noz Caro}]{AA.565.A108.2014}
Giuliano, B., Escribano, R., Martin-Domenech, R., Dartois, E., \& Mu\~noz Caro, G. 2014, Astronomy \& Astrophysics, 565, A108

\bibitem[{Giuliano {et~al.}(2019)Giuliano, Gavdush, M{\"{u}}ller, Zaytsev, Grassi, Ivlev, Palumbo, Baratta, Scir{\`{e}}, Komandin, Yurchenko, \& Caselli}]{AA.629.A112.2019}
Giuliano, B., Gavdush, A., M{\"{u}}ller, B., {et~al.} 2019, \aap, 629, A112

\bibitem[{{Grassi} {et~al.}(2017){Grassi}, {Bovino}, {Haugb{\o}lle}, \& {Schleicher}}]{Grassi2017}
{Grassi}, T., {Bovino}, S., {Haugb{\o}lle}, T., \& {Schleicher}, D.~R.~G. 2017, \mnras, 466, 1259

\bibitem[{Hagen {et~al.}(1981)Hagen, Tielens, \& Greenberg}]{Hagen_CP_1981}
Hagen, W., Tielens, A., \& Greenberg, J. 1981, Chemical Physics, 56, 367

\bibitem[{He {et~al.}(2022)He, Diamant, Wang, Yu, Rocha, Rachid, \& Linnartz}]{He_ApJ_2022}
He, J., Diamant, S., Wang, S., {et~al.} 2022, The Astrophysical Journal, 925, 179

\bibitem[{Hsieh {et~al.}(2013)Hsieh, Campen, Okuno, Backus, Nagata, \& Bonn}]{Hsieh_PNAS_2013}
Hsieh, C.-S., Campen, R.~K., Okuno, M., {et~al.} 2013, Proceedings of the National Academy of Sciences, 110, 18780

\bibitem[{Huang {et~al.}(2013)Huang, Zhang, Ma, Li, Zhou, Zhou, Zheng, \& Sun}]{SR.3.3005.2013}
Huang, Y., Zhang, X., Ma, Z., {et~al.} 2013, Scientific Reports, 3, 3005

\bibitem[{Hudgins {et~al.}(1993)Hudgins, Sandford, Allamandola, \& Tielens}]{Hudgins1993}
Hudgins, D., Sandford, S., Allamandola, L., \& Tielens, A. 1993, \apjs, 86, 713

\bibitem[{{Hudson} {et~al.}(2025){Hudson}, {Yarnall}, \& {Materese}}]{Hudson2025}
{Hudson}, R.~L., {Yarnall}, Y.~Y., \& {Materese}, C.~K. 2025, ACS Earth and Space Chemistry, 9, 2455

\bibitem[{Ipatova {et~al.}(1967)Ipatova, Maradudin, \& Wallis}]{PR.155.3.882.1967}
Ipatova, I., Maradudin, A., \& Wallis, R. 1967, Physical Review, 155, 882

\bibitem[{{Juvela} {et~al.}(2018){Juvela}, {He}, {Pattle}, {Liu}, {Bendo}, {Eden}, {Feh{\'e}r}, {Michel}, {Fuller}, {Hirano}, {Kim}, {Li}, {Liu}, {Malinen}, {Marshall}, {Paradis}, {Parsons}, {Pelkonen}, {Rawlings}, {Ristorcelli}, {Samal}, {Tatematsu}, {Thompson}, {Traficante}, {Wang}, {Ward-Thompson}, {Wu}, {Yi}, \& {Yoo}}]{Juvela2018}
{Juvela}, M., {He}, J., {Pattle}, K., {et~al.} 2018, \aap, 612, A71

\bibitem[{{Kimmel} {et~al.}(2001){Kimmel}, {Dohn{\'a}lek}, {Stevenson}, {Smith}, \& {Kay}}]{Kimmel2001}
{Kimmel}, G.~A., {Dohn{\'a}lek}, Z., {Stevenson}, K.~P., {Smith}, R.~S., \& {Kay}, B.~D. 2001, \jcp, 114, 5295

\bibitem[{Kofman {et~al.}(2019)Kofman, He, Loes~ten Kate, \& Linnartz}]{Kofman_ApJ_2019}
Kofman, V., He, J., Loes~ten Kate, I., \& Linnartz, H. 2019, The Astrophysical Journal, 875, 131

\bibitem[{Kruczkiewicz {et~al.}(2026)Kruczkiewicz, Gavdush, Ribeiro, Campisi, Vyjidak, Giuliano, Komandin, Garnov, Grassi, Theul{\`{e}}, Zaytsev, Ivlev, \& Caselli}]{AA_N2_2025}
Kruczkiewicz, F., Gavdush, A., Ribeiro, F., {et~al.} 2026, \aap, 707, A344

\bibitem[{Leger {et~al.}(1979)Leger, Klein, de~Cheveigne, Guinet, Defourneau, \& Belin}]{AA.79.256.1979}
Leger, A., Klein, J., de~Cheveigne, S., {et~al.} 1979, \aap, 79, 256

\bibitem[{{Li} \& {Draine}(2001)}]{Li_2001}
{Li}, A. \& {Draine}, B.~T. 2001, \apj, 554, 778

\bibitem[{Loerting {et~al.}(2020)Loerting, Fuentes-Landete, Tonauer, \& Gasser}]{Loerting_CC_2020}
Loerting, T., Fuentes-Landete, V., Tonauer, C., \& Gasser, T. 2020, Communications Chemistry, 3, 109

\bibitem[{{Lombardi} {et~al.}(2014){Lombardi}, {Bouy}, {Alves}, \& {Lada}}]{Lombardi_2014}
{Lombardi}, M., {Bouy}, H., {Alves}, J., \& {Lada}, C.~J. 2014, \aap, 566, A45

\bibitem[{Mallamace {et~al.}(2007)Mallamace, Branca, Broccio, Corsaro, Mou, \& Chen}]{PNAS.104.47.18387.2007}
Mallamace, F., Branca, C., Broccio, M., {et~al.} 2007, Proceedings of the National Academy of Sciences, 104, 18387

\bibitem[{Martin(1967)}]{PR.161.1.143.1967}
Martin, P. 1967, Physical Review, 161, 143

\bibitem[{{Mason} {et~al.}(2006){Mason}, {Dawes}, {Holtom}, {Mukerji}, {Davis}, {Sivaraman}, {Kaiser}, {Hoffmann}, \& {Shaw}}]{Mason2006}
{Mason}, N.~J., {Dawes}, A., {Holtom}, P.~D., {et~al.} 2006, Faraday Discussions, 133, 311

\bibitem[{Mastrapa {et~al.}(2009)Mastrapa, Sandford, Roush, Cruikshank, \& Dalle~Ore}]{AJ.701.2.1347.2009}
Mastrapa, R., Sandford, S., Roush, T., Cruikshank, D., \& Dalle~Ore, C. 2009, The Astrophysical Journal, 701, 1347

\bibitem[{{Mathis} {et~al.}(1977){Mathis}, {Rumpl}, \& {Nordsieck}}]{1977ApJ...217..425M}
{Mathis}, J.~S., {Rumpl}, W., \& {Nordsieck}, K.~H. 1977, \apj, 217, 425

\bibitem[{Maty {et~al.}(2021)Maty, Satorre, \& Escibano}]{Satorre_PCCP_2021}
Maty, B., Satorre, M., \& Escibano, R. 2021, Phys. Chem. Chem. Phys., 23, 9532

\bibitem[{Medcraft {et~al.}(2012)Medcraft, McNaughton, Thompson, Appadoo, Bauerecker, \& Robertson}]{AJ.758.1.17.2012}
Medcraft, C., McNaughton, D., Thompson, C., {et~al.} 2012, The Astrophysical Journal, 758, 17

\bibitem[{Medcraft {et~al.}(2013)Medcraft, McNaughton, Thompson, Appadoo, Bauerecker, \& Robertson}]{PCCP.15.10.3630.2013}
Medcraft, C., McNaughton, D., Thompson, C., {et~al.} 2013, Physical Chemistry Chemical Physics, 15, 3630

\bibitem[{{Meg{\'\i}as} {et~al.}(2025){Meg{\'\i}as}, {Jim{\'e}nez-Serra}, {Dulieu}, {Vitorino}, {Mat{\'e}}, {Ciudad}, {Rocha}, {Mart{\'\i}nez Jim{\'e}nez}, \& {Aguirre}}]{Megias2025}
{Meg{\'\i}as}, A., {Jim{\'e}nez-Serra}, I., {Dulieu}, F., {et~al.} 2025, \aap, 702, A87

\bibitem[{{Mifsud} {et~al.}(2022){Mifsud}, {Hailey}, {Herczku}, {Juh{\'a}sz}, {Kov{\'a}cs}, {Sulik}, {Ioppolo}, {Ka{\v{n}}uchov{\'a}}, {McCullough}, {Parip{\'a}s}, \& {Mason}}]{Mifsud2022}
{Mifsud}, D.~V., {Hailey}, P.~A., {Herczku}, P., {et~al.} 2022, European Physical Journal D, 76, 87

\bibitem[{Mishima {et~al.}(1983)Mishima, Klug, \& Whalley}]{JCP.78.11.6399.1983}
Mishima, O., Klug, D., \& Whalley, E. 1983, The Journal of Chemical Physics, 78, 6399

\bibitem[{Moberg {et~al.}(2017)Moberg, Straight, Knight, \& Paesani}]{Moberg_JPCL_2017}
Moberg, D., Straight, S., Knight, C., \& Paesani, F. 2017, The Journal of Physical Chemistry Letters, 8, 2579

\bibitem[{Moore \& Hudson(1992)}]{APJ.401.353.1992}
Moore, M. \& Hudson, R.~L. 1992, \apj, 401, 353

\bibitem[{Moore \& Hudson(1994)}]{AA.103.45.1994}
Moore, M.~H. \& Hudson, R. 1994, Astronomy \& Astrophysics, 103, 45

\bibitem[{Nagasawa {et~al.}(2021)Nagasawa, Sato, Hasegawa, Numadate, Shioya, Shimoaka, Hasegawa, \& Hama}]{Nagasawa2021}
Nagasawa, T., Sato, R., Hasegawa, T., {et~al.} 2021, \apjl, 923, L3

\bibitem[{Narten {et~al.}(1976)Narten, Venkatesh, \& Rice}]{Narten_JCP_1976}
Narten, A., Venkatesh, C., \& Rice, S. 1976, The Journal of Chemical Physics, 64, 1106

\bibitem[{Noble {et~al.}(2024)Noble, Fraser, Smith, Dartois, Boogert, Cuppen, Dickinson, Dulieu, Egami, Erkal, Giuliano, Husquinet, Lamberts, Mat{\'e}, McClure, Palumbo, Shimonishi, Sun, Bergner, Brown, Caselli, Congiu, Drozdovskaya, Herrero, Ioppolo, Jimenez-Serra, Linnartz, Melnick, McGuire, Oberg, Perotti, Qasim, Rocha, \& Urso}]{Noble_NatAst_2024}
Noble, J., Fraser, H., Smith, Z., {et~al.} 2024, Nature Astronomy, 8, 1169

\bibitem[{Nozari {et~al.}(2025)Nozari, Sadavoy, Chapillon, Mason, Friesen, Lowe, Stanke, Di~Francesco, Henning, Zhang, \& Stutz}]{Nozari2025}
Nozari, P., Sadavoy, S., Chapillon, E., {et~al.} 2025, The Astrophysical Journal, 979, 142

\bibitem[{{{\"O}berg} {et~al.}(2011){{\"O}berg}, {Boogert}, {Pontoppidan}, {van den Broek}, {van Dishoeck}, {Bottinelli}, {Blake}, \& {Evans}}]{Oberg_2011}
{{\"O}berg}, K.~I., {Boogert}, A.~C.~A., {Pontoppidan}, K.~M., {et~al.} 2011, \apj, 740, 109

\bibitem[{{Ohashi} {et~al.}(2018){Ohashi}, {Sanhueza}, {Sakai}, {Kandori}, {Choi}, {Hirota}, {Nguyễn-Lu'o'ng}, \& {Tatematsu}}]{Ohashi2018}
{Ohashi}, S., {Sanhueza}, P., {Sakai}, N., {et~al.} 2018, \apj, 856, 147

\bibitem[{{Ormel} {et~al.}(2011){Ormel}, {Min}, {Tielens}, {Dominik}, \& {Paszun}}]{Ormel2011}
{Ormel}, C.~W., {Min}, M., {Tielens}, A.~G.~G.~M., {Dominik}, C., \& {Paszun}, D. 2011, \aap, 532, A43

\bibitem[{{Ossenkopf} \& {Henning}(1994)}]{1994AA.291.943O}
{Ossenkopf}, V. \& {Henning}, T. 1994, Astronomy \& Astrophysics, 291, 943

\bibitem[{Palumbo(2005)}]{Palumbo_2005}
Palumbo, M. 2005, Journal of Physics: Conference Series, 6, 211

\bibitem[{Perotti {et~al.}(2020)Perotti, Rocha, Jørgensen, Kristensen, Fraser, \& Pontoppidan}]{Perotti_AA_2020}
Perotti, G., Rocha, W. R.~M., Jørgensen, J.~K., {et~al.} 2020, \aap, 643, A48

\bibitem[{{Preibisch} {et~al.}(1993){Preibisch}, {Ossenkopf}, {Yorke}, \& {Henning}}]{1993A&A...279..577P}
{Preibisch}, T., {Ossenkopf}, V., {Yorke}, H.~W., \& {Henning}, T. 1993, \aap, 279, 577

\bibitem[{Rocha {et~al.}(2024)Rocha, Gomes~Rachid, McClure, He, \& Linnartz}]{Rocha_AA_2024}
Rocha, W., Gomes~Rachid, M., McClure, M., He, J., \& Linnartz, H. 2024, \aap, 681, A9

\bibitem[{Rocha {et~al.}(2022)Rocha, Gomes~Rachid, Olsthoorn, Dishoeck, McClure, \& Linnartz}]{Rocha_AA_2022}
Rocha, W., Gomes~Rachid, M., Olsthoorn, B., {et~al.} 2022, \aap, 668, A63

\bibitem[{{Rocha} {et~al.}(2021){Rocha}, {Perotti}, {Kristensen}, \& {J{\o}rgensen}}]{Rocha2021}
{Rocha}, W.~R.~M., {Perotti}, G., {Kristensen}, L.~E., \& {J{\o}rgensen}, J.~K. 2021, \aap, 654, A158

\bibitem[{{Sabatini} {et~al.}(2025){Sabatini}, {Bianchi}, {Chandler}, {Cacciapuoti}, {Podio}, {Maureira}, {Codella}, {Ceccarelli}, {Sakai}, {Testi}, {Toci}, {Svoboda}, {Sakai}, {Bouvier}, {Caselli}, {Cuello}, {De Simone}, {J{\'\i}menez-Serra}, {Johnstone}, {Loinard}, {Zhang}, \& {Yamamoto}}]{Sabatini_2025}
{Sabatini}, G., {Bianchi}, E., {Chandler}, C.~J., {et~al.} 2025, \aap, 698, L16

\bibitem[{Salzmann(2019)}]{Salzmann_JCP_2019}
Salzmann, C. 2019, The Journal of Chemical Physics, 150, 060901

\bibitem[{Salzmann \& Murray(2020)}]{NM.19.586.2020}
Salzmann, C. \& Murray, B. 2020, Nature Materials, 19, 586

\bibitem[{Servoin {et~al.}(1980)Servoin, Luspin, \& Gervais}]{PRB.22.11.5501.1980}
Servoin, J., Luspin, Y., \& Gervais, F. 1980, Physical Review B, 22, 5501

\bibitem[{{Stevenson} {et~al.}(1999){Stevenson}, {Kimmel}, {Dohnalek}, {Smith}, \& {Kay}}]{Stevenson1999}
{Stevenson}, K.~P., {Kimmel}, G.~A., {Dohnalek}, Z., {Smith}, R.~S., \& {Kay}, B.~D. 1999, Science, 283, 1505

\bibitem[{Sudera {et~al.}(2020)Sudera, Cyran, Deiseroth, Backus, \& Bonn}]{Sudera_2020}
Sudera, P., Cyran, J., Deiseroth, M., Backus, E., \& Bonn, M. 2020, Journal of the American Chemical Society, 142, 12005

\bibitem[{Tao {et~al.}(2024)Tao, Dai, Moggach, Clode, Fitzgerald, Hodgetts, Harvey, \& Wallace}]{Tao_JCP_2024}
Tao, Y., Dai, X., Moggach, S., {et~al.} 2024, The Journal of Chemical Physics, 160, 214503

\bibitem[{Terada \& Tokunaga(2012)}]{Terada_ApJ_2012}
Terada, H. \& Tokunaga, A. 2012, The Astrophysical Journal, 753, 19

\bibitem[{{Testi} {et~al.}(2014){Testi}, {Birnstiel}, {Ricci}, {Andrews}, {Blum}, {Carpenter}, {Dominik}, {Isella}, {Natta}, {Williams}, \& {Wilner}}]{2014prpl.conf..339T}
{Testi}, L., {Birnstiel}, T., {Ricci}, L., {et~al.} 2014, in Protostars and Planets VI, ed. H.~{Beuther}, R.~S. {Klessen}, C.~P. {Dullemond}, \& T.~{Henning}, 339--361

\bibitem[{Tielens(2013)}]{Tielens_2013}
Tielens, A. 2013, Reviews of Modern Physics, 85, 1021

\bibitem[{Toon {et~al.}(1994)Toon, Tolbert, Koehler, Middlebrook, \& Jordan}]{Toon_JGR_1994}
Toon, O., Tolbert, M., Koehler, B., Middlebrook, A., \& Jordan, J. 1994, \jgr, 99, 25631

\bibitem[{{van Dishoeck} {et~al.}(2013){van Dishoeck}, {Herbst}, \& {Neufeld}}]{vanDishoeck2013}
{van Dishoeck}, E.~F., {Herbst}, E., \& {Neufeld}, D.~A. 2013, Chemical Reviews, 113, 9043

\bibitem[{Walsh {et~al.}(2014)Walsh, Millar, Nomura, Herbst, Widicus~Weaver, Aikawa, Laas, \& Vasyunin}]{Walsh2014}
Walsh, C., Millar, T., Nomura, H., {et~al.} 2014, \aap, 563, A33

\bibitem[{Warren(1984)}]{Warren_AO_1984}
Warren, S. 1984, Applied Optics, 23, 1206

\bibitem[{{Weingartner} \& {Draine}(2001)}]{Weingartner_2001}
{Weingartner}, J.~C. \& {Draine}, B.~T. 2001, \apj, 548, 296

\bibitem[{Westley {et~al.}(1998)Westley, Baratta, \& Baragiola}]{Westley1998}
Westley, M., Baratta, G., \& Baragiola, R. 1998, The Journal of Chemical Physics, 108, 3321

\end{thebibliography}

\begin{appendix}

\section{Parameters of the complex dielectric permittivity model for water ice}
\label{SEC:Appendix}

\begin{table}[!h]
    \caption{Model parameters
    (Eq.~\eqref{EQ:LorentzModel})
    for the Ih \ce{H2O} ice.}
    \small
    \centering 
    \begin{tabular}{>
        {\centering\arraybackslash}m{0.7cm}|>
        {\centering\arraybackslash}m{1.15cm}|>
        {\centering\arraybackslash}m{1.15cm}|>
        {\centering\arraybackslash}m{1.15cm}|>
        {\centering\arraybackslash}m{2.65cm}}
            \rule[-1ex]{0pt}{2.5ex} {$\varepsilon_\infty$}
                                  & {$\Delta\varepsilon_\mathrm{L}$}
                                  & {$\nu_\mathrm{L}$, THz}
                                  & {$\gamma_\mathrm{L}$, THz}
                                  & {Band type}\\
            \hline \hline
            \multirow{20}{*}{$1.735$} & $0.049$ & $1.780$ & $1.000$ & {THz mode} \\ 
                                      \cline{2-5}
                                      & $0.273$ & $4.675$ & $1.049$ &
                                      \multirow{7}{*}{Lattice modes} \\ 
                                      & $0.050$ & $5.636$ & $0.899$ \\ 
                                      & $0.004$ & $5.497$ & $0.130$ \\ 
                                      & $0.120$ & $5.980$ & $1.300$ \\ 
                                      & $0.080$ & $6.200$ & $1.000$ \\ 
                                      & $0.250$ & $6.631$ & $0.562$ \\ 
                                      & $0.070$ & $7.675$ & $1.259$ \\ 
                                      \cline{2-5}
                                      & $4 \times 10^{-7}$ & $15.903$ & $6.968$ &
                                      \multirow{5}{*}{Libration} \\ 
                                      & $0.008$ & $19.776$ & $2.156$ \\ 
                                      & $0.011$ & $22.263$ & $3.147$ \\ 
                                      & $0.183$ & $24.565$ & $3.539$ \\ 
                                      & $0.030$ & $26.946$ & $2.296$ \\ 
                                      \cline{2-5}
                                      & $0.007$ & $38.253$ & $14.662$ &
                                      \multirow{3}{*}{Bending} \\ 
                                      & $0.016$ & $44.751$ & $8.967$ \\ 
                                      & $0.015$ & $49.134$ & $5.745$ \\ 
                                      \cline{2-5}
                                      & $0.007$ & $67.381$ & $6.576$ &
                                      \multirow{1}{*}{\ce{O-H} combination}\\ 
                                      \cline{2-5}
                                      & $0.035$ & $94.339$ & $2.579$ &
                                      \multirow{4}{*}{\ce{O-H} stretching} \\ 
                                      & $0.050$ & $96.523$ & $2.616$ \\ 
                                      & $0.009$ & $98.893$ & $2.964$ \\ 
                                      & $0.009$ & $101.187$ & $3.013$ 
        \end{tabular}
    \label{TAB_Ih_Lorentz}
\end{table}

\begin{table}[!h]
    \caption{Model parameters
    (Eq.~\eqref{EQ:LorentzModel})
    for the Ic \ce{H2O} ice.}
    \small
    \centering 
    \begin{tabular}{>
        {\centering\arraybackslash}m{0.7cm}|>
        {\centering\arraybackslash}m{1.15cm}|>
        {\centering\arraybackslash}m{1.15cm}|>
        {\centering\arraybackslash}m{1.15cm}|>
        {\centering\arraybackslash}m{2.65cm}}
            \rule[-1ex]{0pt}{2.5ex} {$\varepsilon_\infty$}
                                  & {$\Delta\varepsilon_\mathrm{L}$}
                                  & {$\nu_\mathrm{L}$, THz}
                                  & {$\gamma_\mathrm{L}$, THz}
                                  & {Band type}\\
            \hline \hline
            \multirow{25}{*}{$1.621$}  & $0.049$ & $1.780$ & $1.000$ & {THz mode} \\ 
                                      \cline{2-5}
                                       & $0.155$ & $3.900$ & $3.009$ &
                                       \multirow{8}{*}{Lattice modes} \\ 
                                       & $0.292$ & $4.750$ & $0.995$ \\ 
                                       & $0.008$ & $5.598$ & $0.130$ \\ 
                                       & $0.105$ & $5.873$ & $0.722$ \\ 
                                       & $0.104$ & $6.311$ & $0.715$ \\ 
                                       & $0.191$ & $6.737$ & $0.462$ \\ 
                                       & $0.090$ & $7.724$ & $1.412$ \\ 
                                       & $0.010$ & $8.650$ & $1.343$ \\ 
                                       \cline{2-5}
                                       & $3 \times 10^{-7}$ & $15.903$ & $6.968$ &
                                       \multirow{5}{*}{Libration} \\ 
                                       & $0.017$ & $18.577$ & $2.535$ \\ 
                                       & $0.030$ & $21.589$ & $4.017$ \\ 
                                       & $0.262$ & $23.727$ & $4.840$ \\ 
                                       & $0.034$ & $26.264$ & $2.907$ \\ 
                                       \cline{2-5}
                                       & $0.002$ & $40.782$ & $14.111$ &
                                       \multirow{6}{*}{Bending} \\ 
                                       & $0.003$ & $44.202$ & $13.155$ \\ 
                                       & $0.008$ & $44.205$ & $6.591$ \\ 
                                       & $0.006$ & $46.503$ & $5.471$ \\ 
                                       & $0.005$ & $48.709$ & $3.559$ \\ 
                                       & $0.004$ & $50.404$ & $2.120$ \\ 
                                       \cline{2-5}
                                       & $0.003$ & $66.630$ & $3.800$ &
                                       \multirow{2}{*}{\ce{O-H} combination} \\ 
                                       & $0.003$ & $69.850$ & $8.264$ \\ 
                                       \cline{2-5}
                                       & $0.023$ & $95.152$ & $3.277$ &
                                       \multirow{4}{*}{\ce{O-H} stretching} \\ 
                                       & $0.074$ & $97.149$ & $4.035$ \\ 
                                       & $0.016$ & $99.372$ & $4.003$ \\ 
                                       & $0.013$ & $102.039$ & $4.005$ 
        \end{tabular}
    \label{TAB_Ic_Lorentz}
\end{table}

\begin{table}[!h]
    \caption{Model parameters
    (Eq.~\eqref{EQ:LorentzModel})
    for the ASW \ce{H2O} ice.}
    \small
    \centering 
    \begin{tabular}{>
        {\centering\arraybackslash}m{0.7cm}|>
        {\centering\arraybackslash}m{1.15cm}|>
        {\centering\arraybackslash}m{1.15cm}|>
        {\centering\arraybackslash}m{1.15cm}|>
        {\centering\arraybackslash}m{2.65cm}}
            \rule[-1ex]{0pt}{2.5ex} {$\varepsilon_\infty$}
                                  & {$\Delta\varepsilon_\mathrm{L}$}
                                  & {$\nu_\mathrm{L}$, THz}
                                  & {$\gamma_\mathrm{L}$, THz}
                                  & {Band type}\\
            \hline \hline
            \multirow{43}{*}{$1.742$}  & $0.049$ & $1.780$ & $1.000$ & {THz mode} \\ 
                                      \cline{2-5}
                                       & $0.172$ & $3.995$ & $1.116$ &
                                       \multirow{7}{*}{Lattice modes} \\ 
                                       & $0.147$ & $4.831$ & $1.103$ \\ 
                                       & $0.115$ & $5.583$ & $1.086$ \\ 
                                       & $0.082$ & $6.094$ & $0.992$ \\ 
                                       & $0.073$ & $6.575$ & $0.900$ \\ 
                                       & $0.059$ & $7.178$ & $1.076$ \\ 
                                       & $0.039$ & $8.006$ & $1.237$ \\ 
                                       \cline{2-5}
                                       & $0.007$ & $12.757$ & $1.225$ & \multirow{11}{*}{Libration} \\ 
                                       & $0.025$ & $14.270$ & $2.106$ \\ 
                                       & $0.046$ & $15.691$ & $2.621$ \\ 
                                       & $0.015$ & $17.015$ & $1.799$ \\ 
                                       & $0.035$ & $18.205$ & $2.428$ \\ 
                                       & $0.047$ & $19.595$ & $2.427$ \\ 
                                       & $0.073$ & $21.201$ & $2.494$ \\ 
                                       & $0.057$ & $22.661$ & $2.264$ \\ 
                                       & $0.037$ & $23.983$ & $1.982$ \\ 
                                       & $0.022$ & $25.209$ & $1.722$ \\ 
                                       & $0.012$ & $26.370$ & $1.532$ \\ 
                                       \cline{2-5}
                                       & $0.002$ & $40.527$ & $3.711$ &
                                       \multirow{6}{*}{Bending} \\ 
                                       & $0.004$ & $43.514$ & $4.760$ \\ 
                                       & $0.003$ & $46.111$ & $3.695$ \\ 
                                       & $0.002$ & $47.643$ & $2.385$ \\ 
                                       & $0.006$ & $49.468$ & $2.148$ \\ 
                                       & $0.003$ & $50.733$ & $1.652$ \\ 
                                       \cline{2-5}
                                       & $0.0004$ & $63.770$ & $4.640$ &
                                       \multirow{6}{*}{\ce{O-H} combination} \\ 
                                       & $0.002$ & $66.089$ & $4.203$ \\ 
                                       & $0.0007$ & $68.026$ & $5.886$ \\ 
                                       & $2 \times 10^{-5}$ & $69.691$ & $6.176$ \\ 
                                       & $0.0001$ & $70.355$ & $5.193$ \\ 
                                       & $0.0001$ & $71.598$ & $3.291$ \\ 
                                       \cline{2-5}
                                       & $0.002$ & $93.742$ & $1.781$ & \multirow{12}{*}{\ce{O-H} stretching} \\ 
                                       & $0.016$ & $95.566$ & $2.928$ \\ 
                                       & $0.014$ & $97.233$ & $2.738$ \\ 
                                       & $0.017$ & $98.462$ & $2.841$ \\ 
                                       & $0.006$ & $99.686$ & $1.762$ \\ 
                                       & $0.007$ & $100.724$ & $1.837$ \\ 
                                       & $0.005$ & $101.783$ & $1.532$ \\ 
                                       & $0.001$ & $102.419$ & $3.234$ \\ 
                                       & $0.003$ & $102.777$ & $1.390$ \\ 
                                       & $0.002$ & $103.624$ & $1.332$ \\ 
                                       & $0.001$ & $104.511$ & $1.278$ \\ 
                                       & $0.0006$ & $105.586$ & $1.354$ \\ 
                                       \cline{2-5}
                                       & $9 \times 10^{-5}$ & $109.904$ & $0.444$ & \multirow{1}{*}{Dangling \ce{O-H}} 
    \end{tabular}
    \label{TAB_ASW_Lorentz}
\end{table}

\end{appendix}

\end{document}